\pdfoutput=1

\documentclass[10pt,aps,pra,twocolumn,showpacs,superscriptaddress]{revtex4}
\usepackage{amsmath}
\usepackage{latexsym}
\usepackage{amssymb}
\usepackage{graphicx}
\usepackage[colorlinks=true,citecolor=blue, urlcolor=blue]{hyperref}
\usepackage{float}

\begin{document}

\title{Macroscopic locality with equal bias reproduces with high fidelity a quantum distribution achieving the Tsirelson's bound}

\author{Md. Rajjak Gazi}
\email{rajjakgazimath@gmail.com}
\affiliation{Physics and Applied Mathematics Unit, Indian Statistical Institute, 203 B.T. Road, Kolkata-700108, India}

\author{Manik Banik}
\email{manik11ju@gmail.com}
\affiliation{Physics and Applied Mathematics Unit, Indian Statistical Institute, 203 B.T. Road, Kolkata-700108, India}

\author{Subhadipa Das}
\email{sbhdpa.das@bose.res.in}
\affiliation{S.N. Bose National Center for Basic Sciences, Block JD, Sector III, Salt Lake, Kolkata-700098, India}

\author{Ashutosh Rai}
\email{arai@bose.res.in}
\affiliation{S.N. Bose National Center for Basic Sciences, Block JD, Sector III, Salt Lake, Kolkata-700098, India}

\author{Samir Kunkri}
\email{skunkri@yahoo.com}
\affiliation{Mahadevananda Mahavidyalaya, Monirampore, Barrakpore, North 24 Parganas-700120, India}

\begin{abstract}
Two physical principles, Macroscopic Locality (ML) and Information Causality (IC), so far, have been most successful in distinguishing quantum correlations from post-quantum correlations. However, there are also some post-quantum probability distributions which cannot be distinguished with the help of these principles. Thus, it is interesting to see whether consideration of these two principles, separately, along with some additional physically plausible constraints, can explain some interesting quantum features which are otherwise hard to reproduce. In this paper we show that, in a Bell-CHSH scenario, ML along with constraint of \emph{equal-biasness} for the concerned observables, almost reproduces the quantum joint probability distribution corresponding to maximal quantum Bell violation, which is unique up to relabeling. From this example and earlier work of Cavalcanti, Salles and Scarani, we conclude that IC and ML are inequivalent physical principles; satisfying one does not imply that the other is satisfied. 
\end{abstract}
\maketitle
\section{Introduction}
There exist quantum correlations which cannot have any local-realistic description \cite{Einstein,Bell}. Correlations with this surprising property are called nonlocal correlations and are often witnessed through a violation of the celebrated Bell-Clauser-Horne-Shimony-Holt (B-CHSH) inequality \cite{Bell,Clauser} (see also \cite{Brunner} for a review on Bell nonlocality). The violation of B-CHSH inequality in quantum mechanics is restricted by the value $2\sqrt{2}$ (Tsirelson bound) \cite{ Tsirelson}, whereas, for general post-quantum correlations compatible with the no-signaling (NS) principle, B-CHSH expression can achieve the maximum algebraic value $4$ \cite{Popescu}. Deriving all (and only) quantum correlations from a minimal set of physical axioms \cite{Hardy}, and in particular, explaining the typical features of quantum nonlocality is a problem of great research interest in quantum foundations \cite{Pawlowski,Navascues,Oppenheim,Banik}.

Physical principles, like Information Causality (IC) \cite{Pawlowski} and Macroscopic Locality (ML)\cite{Navascues}, have been proposed with an aim to distinguish quantum correlations from the other generalized no-signaling correlations. Interestingly both of these principles can explain the maximum violation of B-CHSH inequality in quantum mechanics. These principles have also been studied in the context of other nonlocal quantum formulations \cite{allcock,ali,gazi,yang}. A comparative study of IC and ML has been made in \cite{Cavalcanti}. The authors \cite{Cavalcanti} have provided certain non quantum correlations which satisfy ML but violate IC. This seems to imply that in distinguishing
quantum correlations from post-quantum correlations, IC is better than ML. In this work, we provide a situation where ML is a stronger condition than IC. This occurs at the Tsirelson bound (CHSH violation $= 2 \sqrt{2}$) when one imposes the condition of equal-biasness (described in the next paragraph).

In quantum mechanics, quantum correlations which lead to the Tsirelson bound for a B-CHSH expression have two notable features; (a) The outcomes of any local observable are completely random and (b) the joint probability distribution is unique up to some relabeling.
Neither IC nor ML can reproduce these features by its own. Moreover, it follows from the result of Colbeck and Renner \cite{colren} that, for any no signaling model which can simulate statistics of a maximally entangled state (and in particular, a joint probability distribution giving Tirelson bound), local outcomes must be \emph {equally biased}. On the other hand, IC or ML do not say anything about the feature of 
equal biasness inherent to local statistics of a maximally entangled state. Hence, along with IC or ML we also incorporate the {\it condition of equal-biasness}, which restricts the probability distributions for the outcomes of all the four local measurements involved to be identical. We show that powered by this extra condition, ML almost reproduces all the features of a quantum joint probability distribution which achieves the Tsirelson bound. However, this is not the case with IC which stands very far in this context.

We introduce a measure for the biasness of a measurement with binary outcomes. For such a measurement if the two outcomes occur with the probability $\alpha$ and $1-\alpha$, the biasness (in  percentage) is defined as $|1-2\alpha|\times 100$. We find that at the most $30\%$ biasness can be assigned to each of the four local measurements in a generalized no-signaling theory having Bell-violation $2\sqrt{2}$, if IC condition is used. On the other hand, the corresponding biasness under ML turns out to be $0.04\%$. It is note worthy here that in quantum theory, due to complete randomness of the local measurements, this biasness is zero. Interestingly, not only the local  measurement biasness but the full joint probability distribution under ML turns out to be almost the same as the quantum joint distribution (cp. TABLE-\ref{table0}, TABLE-\ref{table4}).

This paper is organized as as follows. In section(II) we find the maximum equal-biasness for all the four observables (in a B-CHSH scenario) under the no-signaling (NS) condition. Next, in section(III), under the information causality (IC) principle maximum equal-biasness for all the four observables is derived, and the resulting distribution is compared with the quantum mechanical distribution. Then, in section(IV), we invoke the macroscopic locality (ML) principle to obtain the maximum equal-biasness and the resulting joint probability distribution, and compare our results with those obtained from quantum mechanical consideration. Finally, in section(V) we give our conclusions.

\begin{center}
\begin{table}[t]
\begin{tabular}{|c|c|c|c|c|}
		\hline
	$xy\backslash ab$ &  $0_A0_B$    &   $0_A1_B$  & $1_A0_B$   & $1_A1_B$  \\
		\hline\hline
	$\mathbf{0_A0_B}$  & $0.427$& $0.073$& $0.073$ & $0.427$ \\
	\hline
	$\mathbf{0_A1_B}$  & $0.427$& $0.073$& $0.073$ & $0.427$ \\
	\hline
	$\mathbf{1_A0_B}$  & $0.427$& $0.073$& $0.073$ & $0.427$ \\
	\hline
	$\mathbf{1_A1_B}$  & $0.073$& $0.427$& $0.427$ & $0.073$ \\
	\hline
\end{tabular}
\caption{Quantum probability distribution with B-CHSH value $2\sqrt{2}$. Two different spin measurements are denoted by `{\bf $0$}' and `{\bf $1$}' and the subscripts `$A$' and `$B$' is used to indicate corresponding inputs and outputs for Alice and Bob respectively. In the table we approximate $\frac{1}{4}(1+\frac{1}{\sqrt{2}})\simeq 0.427$ and $\frac{1}{4}(1-\frac{1}{\sqrt{2}})\simeq 0.073$}\label{table0}
\end{table}
\end{center}
\section{Maximum biasness under NS}\label{ns}
Consider two spatially separated observers, say Alice and Bob, who perform local measurements denoted by $x$ and $y$, respectively, with corresponding binary outcomes $a$ and $b$, say, $x,y,a,b\in\{0,1\}$. $P(ab|xy)$ denotes the probability of obtaining outcome $a$ at Alice's end and outcome $b$ at Bob's end conditioned upon measurement $x$ and $y$ are performed by Alice and Bob, respectively. The full probability distribution $P(ab|xy)$ is characterized by $16$ joint probabilities satisfying following constraints:
\begin{eqnarray}
P(ab|xy)\ge 0;~~\forall~x,y,a,b~~\mbox{(positivity)}\\
\sum_{a,b=0}^{1}P(ab|xy)=1;~~\forall~x,y~~\mbox{(normalization)}
\end{eqnarray}
The no-signaling condition implies that from the marginal probability distribution of her measurement Alice cannot get any information about measurement choices at Bob's end and vice-verse. This constraint can be expressed by the following mathematical restrictions on the joint probability distributions:
\begin{eqnarray}
\sum_{b=0}^{1}P(ab|xy)=\sum_{b=0}^{1}P(ab|xy');~~\forall~y\neq y',x,a\\
\sum_{a=0}^{1}P(ab|xy)=\sum_{a=0}^{1}P(ab|x'y);~~\forall~x\neq x',y,b
\end{eqnarray}
$y$ and $y'$ denote two different measurement choices at Bob's end and similarly different measurement choices at Alice's end are denoted by $x$ and $x'$. Due to normalization and no-signaling condition, out of $16$ joint probabilities only $8$ are independent and they form a eight-dimensional polytope structure \cite{Barrett2}.
\begin{center}
\begin{table}[b]
\begin{tabular}{|c|c|c|c|c|}
	\hline
$xy\backslash ab$ &  $0_A0_B$    &   $0_A1_B$  & $1_A0_B$   & $1_A1_B$  \\
	\hline\hline
$\mathbf{0_A0_B}$  & $c_1$& $m_1-c_1$& $n_1-c_1$ & $1+c_1-m_1-n_1$ \\
\hline
$\mathbf{0_A1_B}$   & $c_2$& $m_1-c_2$& $n_2-c_2$ & $1+c_2-m_1-n_2$\\
\hline
$\mathbf{1_A0_B}$   & $c_3$& $m_2-c_3$& $n_1-c_3$ & $1+c_3-m_2-n_1$\\
\hline
$\mathbf{1_A1_B}$  & $c_4$& $m_2-c_4$& $n_2-c_4$ & $1+c_4-m_2-n_2$\\
\hline
\end{tabular}
\caption{Bipartite two input-two output no-signaling probability distribution.}\label{table1}
\end{table}
\end{center}
Set of local correlations and quantum correlations are strictly contained within the no-signaling polytope. Set of local correlation itself forms a convex polytope whereas the set of quantum correlations is convex but is not a polytope as the number of extremal points is not finite \cite{Pitowsky,Werner,Landau}. To date no finite set of physical conditions can identify the set quantum correlations uniquely. For input $x$ at Alice's end and input $y$ at Bob's end we can write the no-signaling joint distribution as $P(ab|xy)\equiv (c,m-c,n-c,1+c-m-n)$ (see TABLE-\ref{table1}), where $P(a|x)\equiv(m,1-m)$ and $P(b|y)\equiv(n,1-n)$ are marginal distribution for Alice and Bob, respectively. Positivity is guaranteed by the condition $\mbox{max}\{0,m+n-1\}\le c\le\mbox{min}\{m,n\}$, for each pair of $x$ and $y$. The B-CHSH quantity for the correlation in TABLE-\ref{table1} is given by:
\begin{eqnarray*}
\mbox{B-CHSH}&=&|\langle\mathbf{0_A0_B}\rangle+\langle\mathbf{0_A1_B}\rangle+\langle\mathbf{1_A0_B}\rangle-\langle\mathbf{1_A1_B}\rangle|\\
&=&|2+4(c_1+c_2+c_3-c_4)-4(m_1+n_1)|
\end{eqnarray*}
$\langle *\rangle$ denotes expectation value. It is well known that for the correlations satisfying the NS, the B-CHSH quantity can take any value up to $4$, whereas in quantum mechanics the maximum value of the B-CHSH expression is $2\sqrt{2}$ (Tsirelson bound) \cite{Tsirelson}. In \cite{Hall}, Hall showed that within  generalized no-signaling theory, for a correlation with B-CHSH inequality violation equal to Tsirelson bound at the most $60\%$ biasness can be assigned to any of the four observables. Interestingly we find that on imposing \emph{equal-biasness} condition along with fixing the B-CHSH value to the Tsirelson bound, maximum $60\%$ biasness can be assigned to all the four observables. Under the equal-biasness condition, i.e. $m_1=m_2=n_1=n_2=p$ (say), the probability distribution in TABLE-\ref{table1} modifies such that the B-CHSH expression is $2+4(c_1+c_2+c_3-c_4)-8p$. Then, for the B-CHSH value $2\sqrt{2}$ the value of marginal probability for which the local outcomes are maximally biased is given by $p=\frac{3-\sqrt{2}}{2}\approx 0.8$, which implies that the maximum equal-biasness for all the four observable that can be achieved under NS is $|1-2p|\times 100\% \approx 60\%$ which is significantly different from the quantum mechanical zero biasness.
\section{Maximum biasness under IC}\label{ic}
The information causality (IC) principle \cite{Pawlowski} is a generalization of no-signaling condition and can be formulated quantitatively through an information processing game played between two parties, say Alice and Bob. Alice receives a randomly generated $N$-bit string $\vec{x}=(x_0,x_1,...,x_{N-1})$, and Bob is asked to guess Alice's $i$-th bit where $i$ is randomly chosen from the set $\{0,1,2,...,N-1\}$. Alice is allowed to send a $M$-bit message ($M<N$). Alice and Bob can pre-share no-signaling resources (correlations) which they exploit according to some pre-agreed strategy while playing this game. Let Bob's answer be denoted by $\beta_{i}$. Then, the information that Bob can potentially acquire, about the variable $x_i$ of Alice, is given by the Shannon mutual information $I(x_i:\beta_i)$. The statement of the IC is that the total potential information \cite{Pawlowski} about Alice's bit string $\vec{x}$ accessible to Bob cannot exceed the volume of message he received from Alice, i.e.,
\begin{equation}
\mathbb{I}=\sum^{N}_{i=1} I(x_i:\beta_i)\leq M
\end{equation}
Then the principle of information causality says that physically allowed theories must have $I \le M$. Both classical and quantum correlations have been proved to satisfy the IC principle\cite{Pawlowski}. It was further shown that, if Alice and Bob share arbitrary two input and two output no-signaling correlations corresponding to conditional probabilities $P(ab|xy)$, then by applying a protocol by Van Dam \cite{Van} and Wolf and Wullschleger \cite{wolf}, one can derive a necessary condition for respecting the IC principle. The mathematical form of necessary conditions for information causality from Alice to Bob (A$\rightarrow$B) can be expressed as:
\begin{equation}\label{ic1}
 E_1^2+E_2^2\leq 1
\end{equation}
where $E_i=2P_i^A-1$ for $i=1,2$ and $P_1^A=\frac{1}{2}[P(a=b|00)+P(a=b|10)]$ and $P_2^A=\frac{1}{2}[P(a=b|01)+P(a\neq b|11)]$. Similarly the necessary conditions for information causality from Bob to Alice (B$\rightarrow$A) can be expressed as:
\begin{equation}\label{ic2}
 F_1^2+F_2^2\leq 1
\end{equation}
where $F_i=2P_i^B-1$ for $i=1,2$ and $P_1^B=\frac{1}{2}[P(a=b|00)+P(a=b|01)]$ and $P_2^B=\frac{1}{2}[P(a=b|10)+P(a\neq b|11)]$.

Applications of the IC principle in the study of both bi-partite and tri-partite correlations have produced some interesting results \cite{allcock,ali,gazi,yang}. Here we find the maximum probability that can be assigned to the local observables of a general $2\sqrt{2}$ correlations under the IC principle and the \emph{equal-biasness} condition. With the \emph{equal-biasness} constraint (i.e. $m_1=m_2=n_1=n_2=p$) Eq.(\ref{ic1}) and Eq.(\ref{ic2}) becomes:
\begin{eqnarray}\label{condi1}
(1-4p)^2+4(c_1+c_3)^2+4(c_2-c_4)^2 \nonumber\\
+4(c_1+c_3)(1-4p)\leq 1
\end{eqnarray}
\begin{eqnarray}\label{condi2}
(1-4p)^2+4(c_1+c_2)^2+4(c_3-c_4)^2 \nonumber \\
+4(c_1+c_2)(1-4p)\leq 1
\end{eqnarray}
Maximum value of $p$ under the restrictions of Eq.(\ref{condi1}) and Eq.(\ref{condi2}) becomes $p= 0.646469\approx 0.65$, therefore maximum allowed biasness to all the four observables is $|1-2p|\times 100\%\approx 30\%$. Note that this result differs from quantum result (i.e. zero biasness) but less than the corresponding NS value (i.e. $60\%$ biasness). The $16$ probability distribution corresponding to the maximum allowed biasness under IC is given in TABLE-\ref{table3} which is again different from the probability distribution given in TABLE-\ref{table0} or any distribution obtained by relabeling measurements and outcomes in Table-\ref{table0}.
\begin{center}
\begin{table}[h!]
\begin{tabular}{|c|c|c|c|c|}
	\hline
$xy\backslash ab$ &  $0_A0_B$    &   $0_A1_B$  & $1_A0_B$   & $1_A1_B$  \\
	\hline\hline
$\mathbf{0_A0_B}$  & $0.500$& $0.146$& $0.146$ & $0.207$ \\
\hline
$\mathbf{0_A1_B}$  & $0.646$& $0$& $0$ & $0.354$ \\
\hline
$\mathbf{1_A0_B}$  & $0.646$& $0.000$& $0.000$ & $0.354$ \\
\hline
$\mathbf{1_A1_B}$  & $0.293$& $0.354$& $0.354$ & $0.000$ \\
\hline
\end{tabular}
\caption{In worst case (i.e., maximally biased observables), no signaling probability distribution with the B-CHSH value $2\sqrt{2}$ which satisfies necessary condition for respecting the IC is very different from the unique quantum distribution shown in TABLE-\ref{table0}.}\label{table3}
\end{table}
\end{center}
\section{Maximum biasness under ML}\label{ml}
The principle of Macroscopic Locality \cite{Navascues} states that in the limit of large number, say $N$, of correlated particles emitted from a common source, statistics generated from any coarse grained Bell-type experiment with the help of physically realizable sources must be local, i.e, it must satisfy any Bell-type inequality. All quantum correlations have been shown to respect the ML principle. Interestingly, a large number of no-signaling post-quantum correlation fail to satisfy the ML principle; for instance, PR-correlation can be easily shown to violate the ML principle.
\begin{center}
\begin{table}[b]
\begin{tabular}{|c|c|c|c|c|}
	\hline
$xy\backslash ab$ &  $0_A0_B$    &   $0_A1_B$  & $1_A0_B$   & $1_A1_B$  \\
	\hline\hline
$\mathbf{0_A0_B}$  & $0.427$& $0.0737$& $0.0737$ & $0.427$ \\
\hline
$\mathbf{0_A1_B}$  & $0.427$& $0.0737$& $0.0737$ & $0.427$ \\
\hline
$\mathbf{1_A0_B}$  & $0.427$& $0.0737$& $0.0737$ & $0.427$ \\
\hline
$\mathbf{1_A1_B}$  & $0.073$& $0.427$& $0.427$ & $0.073$ \\
\hline
\end{tabular}
\caption{Even in worst case (i.e., maximally biased observables), no signaling probability distribution with the B-CHSH value $2\sqrt{2}$ which satisfies the ML condition is almost same as the unique quantum distribution shown in TABLE-\ref{table0}. Up-to three decimal places the two distributions are in fact identical.}\label{table4}
\end{table}
\end{center}

Consider a situation with a bipartite system where a source emits pairs of particles to Alice and Bob. Suppose, Alice can choose a measurement $x\in\{0,1\}$ from a set of two possible settings, each producing two possible outcomes, denoted as $a\in\{0,1\}$. Similarly, Bob can choose a measurement $y\in \{0,1\}$, with two outcomes $b \in \{0,1\}$. In a microscopic experiment single-particle detections are possible and the collected statistics can be used for estimating $P(a,b|x,y)$. However, contrary to a microscopic experiment, in a macroscopic (coarse grained) experiment in each run $N$ pairs are sent; Alice measures all her particles with setting $x$ and records the number $n_a$ of particles that produced the outcome $a\in\{0,1\}$; Bob does similarly. After repeating this procedure several times, from the coarse grained statistics now one can estimate $P(\vec{n}_A,\vec{n}_B|x,y)$, where $\vec{n}_A = (n_0,n_1)$ and $\vec{n}_B = (n_0,n_1)$. For $N>1$ since Alice and Bob do not know which of their outcomes came from the same pair, some microscopic information is lost. Then, in the limit $N\rightarrow \infty$, the principle of macroscopic locality demands that the macroscopic statistics $P(\vec{n}_A,\vec{n}_B|x,y)$ should not violate any Bell inequality.

On the other hand, in bipartite systems with two dichotomic measurements per site, there exist a hierarchy of Semi-Definite-Programs (SDPs) that eventually converge to the quantum set \cite{Navascues1,Navascues2}. It has been shown that the first step in this hierarchy generates a set of correlations $Q_1$ which exactly coincides the set correlations respecting the ML principle. However, the set $Q_1$ strictly contains the quantum set $Q$ thus showing insufficiency of the ML principle in distinguishing all post-quantum correlations. In what follows, for the bipartite case with two dichotomic observables at each site, we apply the necessary and sufficient criteria for respecting Macroscopic Locality \cite{Navascues1,Navascues2} given by the condition:
\begin{equation}\label{ml1}
|\sum_{x,y=0}^{1}(-1)^{xy}\sin^{-1}(D_{xy})|\leq \pi
\end{equation}
where $D_{xy}=\frac{C_{xy}-C_{x}.C_{y}}{\sqrt{(1-C_{x}^2)(1-C_{y}^2)}}$; with $C_{xy}$, $C_{x}$ and $C_{y}$ defined as:
\begin{eqnarray}
C_{xy}=\sum_{a=b}P(ab|xy)-\sum_{a\neq b}P(ab|xy)\nonumber\\
C_{x}=\sum_{b}[P(0b|x0)-P(1b|x0)]\nonumber\\
C_{y}=\sum_{a}[P(a0|0y)-P(a1|0y)]\nonumber
\end{eqnarray}
where $x,y\in\{0,1\}$. For the distribution given in TABLE-\ref{table1} with condition of \emph{equal-biasness} (i.e. $m_1=m_2=n_1=n_2=p$) the Eq.(\ref{ml1}) becomes:
\begin{eqnarray}\label{ml2}
\left|\sin^{-1} \left(\frac{c_1-p^2}{p(1-p)}\right)  + \sin^{-1} \left( \frac{c_2-p^2}{p(1-p)}\right)\right. \nonumber\\
\left. + \sin^{-1} \left( \frac{c_3-p^2}{p(1-p)}\right)  - \sin^{-1} \left( \frac{c_4-p^2}{p(1-p)}\right)\right| \leq \pi .
\end{eqnarray}
The maximum value of $p$ under the restrictions of Eq.(\ref{ml2}) is strictly larger than $1/2$, as was already noted in \cite{Navascues}. Numerical calculation shows that the maximum value is $p = 0.500226 \simeq 0.5002$. Therefore for a $2\sqrt{2}$ correlation the ML allows at most $0.04\%$ biasness to all the four observables and in comparison to the IC this value is quite closer to the corresponding quantum value. For the maximum allowed biasness under the ML, the probability distribution is given in TABLE-\ref{table4}, which is identical (up-to three decimal places) to quantum distribution given in TABLE-\ref{table0}.
\section{Conclusion}\label{conclusion}
Information Causality and Macroscopic Locality have been proposed to distinguish physical correlations from non physical ones. Though the Tsirelson bound has been derived from separate considerations of IC and ML, but there are probability distributions giving B-CHSH violation equal to the Tsirelson bound $2\sqrt{2}$ which can largely differ from corresponding unique quantum distribution (TABLE-\ref{table0}). However, by imposing equal-biasness condition over the set of bipartite no-signaling probability distributions with the B-CHSH value equal to $2\sqrt{2}$, we show that ML allows negligibly less biasness compared to IC (Please note that corresponding quantum probability distributions have no biasness). Moreover, the probability distributions allowed by ML along with the equal-biasness condition go quite close to the quantum distribution which at the Tsirelson bound is unique up-to relabeling of inputs and outputs. In this respect, the necessary condition for respecting IC allows probability distributions which can largely differ from quantum ones and thus also violates the ML. It would be interesting to explore further whether some other condition along with IC or ML can exactly determine the quantum distribution corresponding to the Tsirelson bound. Further, in view of the result in \cite{Cavalcanti} where Cavalcanti \emph{et al.} have shown that there exists no-signaling probability distributions which satisfy ML but violate IC placing more confidence in IC as a physical principle defining the set of the possible correlations allowed in Nature. Here we have shown a converse, namely we exhibit no-signaling probability distributions that satisfies IC but violate ML. Thus, we conclude that IC and ML are inequivalent physical principles; satisfying one does not necessarily imply that the other is satisfied.

\section*{Acknowledgment}
It is a pleasure to thank Guruprasad Kar for many stimulating discussion. RG and MB acknowledge S. K. Choudhary for fruitful discussions. SD and AR acknowledges support from the DST project SR/S2/PU-16/2007. AR and SK acknowledge Sibasish Ghosh for fruitful discussions during their recent visit to IMSc, Chennai.

\end{document}